\documentclass[a4paper]{article}

\usepackage{INTERSPEECH2022}
\usepackage{svg}
\usepackage{subcaption}

\title{Introducing ECAPA-TDNN and Wav2Vec2.0 Embeddings to Stuttering Detection}

\name{Shakeel A.~Sheikh$^1$, Md Sahidullah$^1$, Fabrice Hirsch$^2$, Slim Ouni$^1$}

\address{
  $^1$Universit\'{e} de Lorraine, CNRS, Inria, LORIA, F-54000, Nancy, France\\
 $^2$Universit\'{e} Paul-Val\'{e}ry Montpellier, CNRS, Praxiling, Montpellier, France}
 
\email{\{shakeel-ahmad.sheikh, md.sahidullah, slim.ouni\}@loria.fr, fabrice.hirsch@univ-montp3.fr}

\begin{document}

\maketitle
\begin{abstract}
 
The adoption of advanced deep learning (DL) architecture in stuttering detection (SD) tasks is challenging due to the limited size of the available datasets. To this end, this work introduces the application of speech embeddings extracted with pre-trained deep models trained on massive audio datasets for different tasks. In particular, we explore audio representations obtained using emphasized channel attention, propagation, and aggregation-time-delay neural network (ECAPA-TDNN) and Wav2Vec2.0 model trained on VoxCeleb and LibriSpeech datasets respectively. After extracting the embeddings, we benchmark with several traditional classifiers, such as a k-nearest neighbor, Gaussian naive Bayes, and neural network, for the stuttering detection tasks. In comparison to the standard SD system trained only on the limited SEP-28k dataset, we obtain a relative improvement of 16.74\% in terms of overall accuracy over baseline. Finally, we have shown that combining two embeddings and concatenating multiple layers of Wav2Vec2.0 can further improve SD performance up to 1\% and 2.64\% respectively.
\end{abstract}

\vspace{-0.1cm}
\section{Introduction}
Stuttering,~a neuro developmental speech disorder, caused by the failure of speech sensorimotors, is defined by the disturbance of uncontrolled utterances: interjections, and \emph{core behaviours}: blocks, repetitions, and prolongations~\cite{smith2017stuttering, guitar2013stuttering}. Studies show that persons who stutter (PWS) encounter several hardships in social and professional interactions.
 
In addition, more people are progressively interacting with voice assistants, but they ignore and fail to recognize stuttered speech~\cite{sheikh2021machine}, and the stuttering detection (SD) can be exploited to improve automatic speech recognition (ASR) for PWS to access voice assistants such as Alexa, Siri, etc.
\par

Usually, SD is addressed by various listening and brain scan tests~\cite{Ingham1996FunctionallesionIO, smith2017stuttering, sheikh2021machine}. However, this method of SD is high-priced and requires a demanding effort from speech therapists. The presence of uncontrolled utterances are reflecting in the acoustic domain, which helps to discriminate them in various stuttering types. Based on the acoustic cues present in stuttered speech, several people employed machine learning paradigm for SD. Some of the current state-of-the-art SD DL modelling techniques include:~ResNet+BiLSTM~\cite{tedd, melanie}, FluentNet~\cite{fluentnet}, StutterNet~\cite{stutternet}. T.Kourkounakis~\emph{et al}~\cite{fluentnet, tedd} approached the SD as a multiple binary classification problem and trained separate ResNet+BiLSTM classifiers and FluentNet classifiers for each stuttering type. The models were trained using spectrogram input features on a small set of 24 UCLASS speakers. Shakeel~\emph{et al}~\cite{stutternet} approached SD via single branch StutterNet and proposed the first multi-class classifier for SD and its types. The classifier is trained with 20 MFCC input features on a large set of more than 100 UCLASS speakers. Lee~\emph{et al}~\cite{sep28k} recently introduced a new SEP-28k stuttering dataset. Melanie~\emph{et al}~\cite{melanie} exploited phoneme features proposed BiLSTM method for SD. The method is trained on mel-spectral and phoneme based input features by mixing SEP-28k, UCLASS, and FluencyBank datasets. A comprehensive detail on SD methods can be found in the review papers by L.Barrett~\emph{et al}~\cite{sysreview} and Shakeel~\emph{et al}~\cite{sheikh2021machine}. 
\par 
Over the years, several stuttering datasets including: SEP-28k~\cite{sep28k}, UCLASS~\cite{sheikh2021machine}, LibriStutter~\cite{fluentnet}, and FluencyBank~\cite{sep28k} have been developed for investigating different SD models. Even though the SD methods discussed show promising results on these datasets, however,~these datasets are relatively very small and are limited to only a certain number of speakers. Due to the varying nature of stuttering from person-to-person, these small datasets have an inclination to be biased towards these small pool of speakers~\cite{guitar2013stuttering}. In addition, the DL have shown tremendous improvement in ASR~\cite{speechrecog}, emotion detection~\cite{edwav2vec2,speechemotion}, speaker verification~\cite{haqiadversarial}, etc, however the improvement in SD is bounded, most likely due to the limited size of stuttering datasets, which are unable to capture different speaking styles, accents, linguistic content, etc. In addition, collecting medical data requires big-budget and is very taxing, and stuttering data collection is no exception. 
\par 
To tackle this, we use the transfer learning (TL) paradigm that has been successfully used in ASR~\cite{wav2vec2} and emotion recognition~\cite{edwav2vec2}, for instance. In this paper, we mainly focus on the speaker and Wave2Vec2.0 contextual embeddings. Pre-training a model on such a massive datasets can successfully capture the variable speaking styles, emotion behaviours which are extremely important in the SD~\cite{sheikh2021machine}. 

\par
Our primary contributions in this paper are:
\begin{itemize} 
    \item  \vspace{-0.1cm} We explore the use of speaker embeddings extracted from ECAPA-TDNN~\cite{ecappatdnn}, and Wav2Vec2.0 speech representations  for SD, and to the best of our knowledge, is the first ever study in the domain of SD.
    
    \item We provide a novel way for SD, which exploits the information from the fully connected (FC) layer of ECAPA-TDNN~\cite{ecappatdnn} and draws on speech information from several hidden layers of the Wav2vec2.0 model~\cite{wav2vec2}.
    \item We also provide an analysis on the impact of using different layers from Wav2Vec2.0 and their concatenation in SD and also investigate the impact of combining information from ECAPA-TDNN and Wav2Vec2.0 embeddings via score fusion.

\end{itemize}
\vspace{-0.3cm}

\section{Speaker and contextual embeddings}

\subsection{Speaker embeddings}
\begin{figure}
\vspace{-0.7cm}
    \centering
    \includegraphics[scale=0.5]{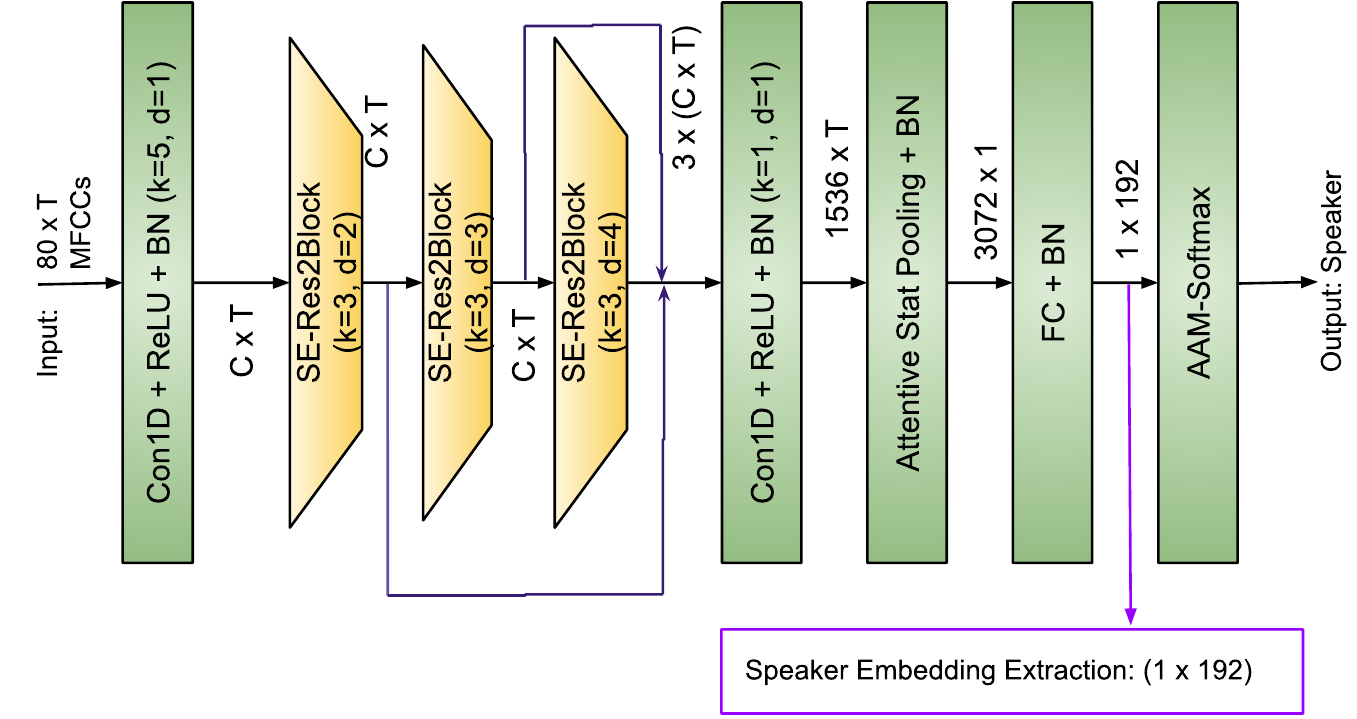} 
    \caption{\small ECAPA-TDNN Block diagram, k is kernel size, d is dilation, C is channel and T is temporal dimension.}
    \vspace{-0.5cm}
    \label{fig:ecapatdnn} 
\end{figure}

 \vspace{-0.15cm}
Speaker embeddings are usually computed from trained neural networks to identify and classify speakers from a group of speakers~\cite{ecappatdnn}. The pre-trained speaker embeddings have been successfully applied in speaker diarization~\cite{ectdnnspkrdiar}. The ECAPA-TDNN is the SOTA for extracing speaker embeddings. As depicted in Fig.~\ref{fig:ecapatdnn}, the ECAPA-TDNN is composed of 1D convolution followed by three 1D squeeze and excitation (SE) Res2Blocks, 1D convolution, attentive statistical pooling, and a FC layer. A non linear ReLU activation and batch normalization (BN) is applied after each layer in SE-Res2Block. The model is fed with 80 dimensional mean normalized log Mel filterbank energies and is trained on a large Voxceleb dataset with AAM-softmax loss function using Adam optimizer having a cycling learning rate between 1e-8 and 1e-3. In this paper, we use ECAPA-TDNN as a feature vector on SEP-28k in order to exploit it for SD. We extract $\mathcal{R}^{1\times192}$ speaker embedding feature vector from the FC layer of ECAPA-TDNN as shown by purple block in Fig.~\ref{fig:ecapatdnn}. 
\vspace{-0.2cm}
\subsection{Wav2Vec2.0 contextual embeddings}
\begin{figure}
\vspace{-0.7cm}
    \centering
    \includegraphics[scale=0.4]{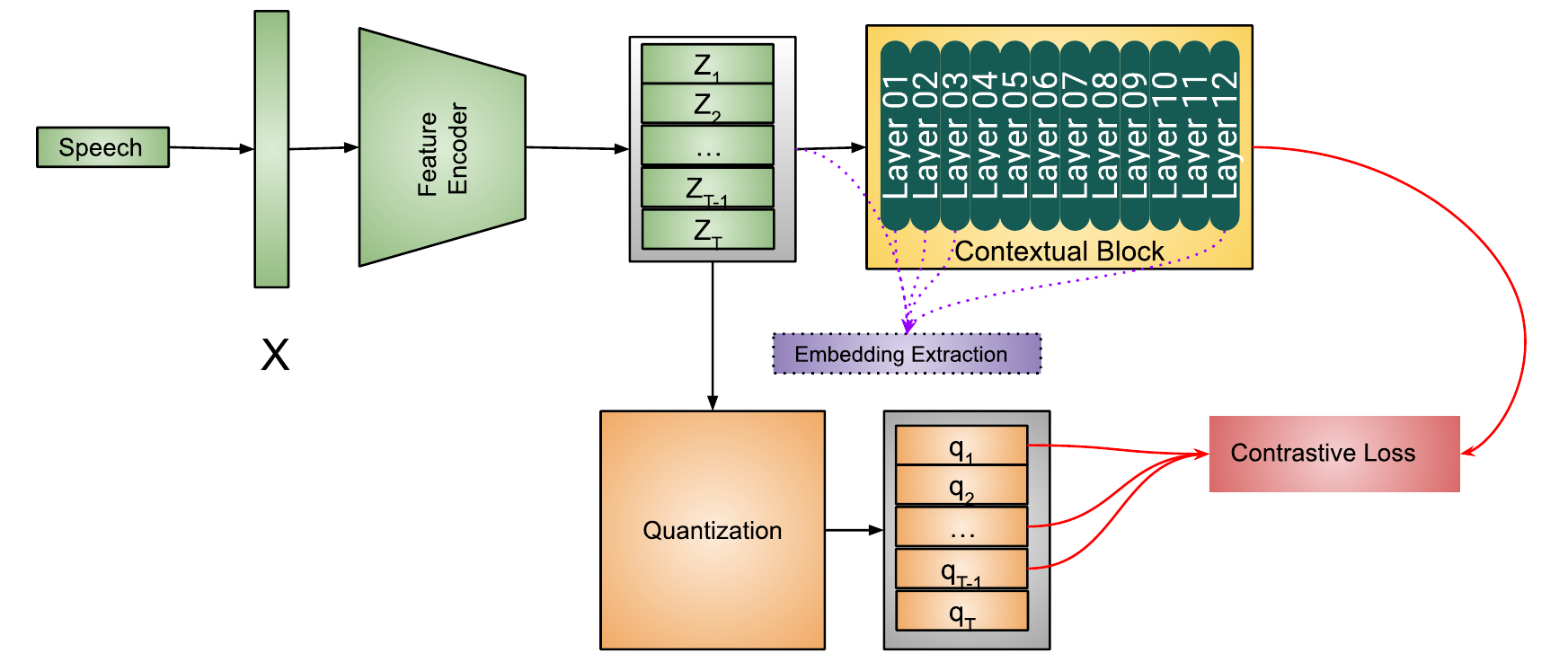} 
    \caption{Block diagram for Wav2Vec2.0 architecture.} \vspace{-0.3cm}
    \label{fig:wa2vec2.0} 
\end{figure}

 \vspace{-0.15cm}
The WaveVec2.0 model is a self-supervised representation learning framework of raw audio, and is comprised of three modules including feature encoder $f\colon \mathcal{X}\mapsto \mathcal{Z}$, contextual block transformer $g\colon \mathcal{Z}\mapsto \mathcal{C}$ and quantization block $\mathcal{Z}\mapsto \mathcal{Q}$ as depicted in Fig.~\ref{fig:wa2vec2.0}. The feature encoder is comprised of multi-layer 1D convolution blocks followed by BN and GELU activation functions, takes the normalized raw input $\mathcal{X}$ and encodes it into local feature representations~$\mathcal{Z} = f(\mathcal{X})$. The authors have released two pre-trained feature embeddings with dimensions of $768$ (base) and $1024$ (large) and we are using the base one in our study. These encoded feature representations of size $\mathcal{Z}^{T\times 768}$ are then fed to contextual transformer block  to learn contextual speech representations $\mathcal{C} = g(\mathcal{Z})$. The paper uses two different transformer 
networks with the base model consisting of 12 blocks having eight attention heads at each block, and the large model is comprised of 24 blocks with 16 attention heads at each block. The feature representations $\mathcal{Z}$ are also fed to the quantization module which is comprised of two codebooks having 320 possible entries in each. For each vector representation $\textit{z}_i$ $\in \mathcal{Z}$, a logit of $\mathcal{R}^{2\times320}$ is chosen using~(\ref{gumbel}) by concatenating the corresponding entries from each codebook, which is then followed by linear transformation to produce the quantized vector $\textit{q}_i$ of the local feature encoder representation $\textit{z}_i$ $\in \mathcal{Z}$. 
\begin{equation}
    p_{g,v} = \frac{\exp(l_{g,v} + \eta_v)/\tau}{\sum_{k=1}^{V}\exp(l_{g,v} + \eta_v)/\tau}
    \label{gumbel}
\end{equation}
where $l$ is logit, $v$ is v-th codebook entry, $g$ is codebook group, $\eta=-\log(-log(u))$ with $u$ are uniform samples from $\mathcal{U}(0,1)$, and $\tau$ is the temperature which controls the randomness. 
\par 
Similar to the masked language modelling,~the model is pre-trained in a self-supervised fashion using eq.~(\ref{eq:contrastiveloss}) by randomly masking certain time stamp representation vectors of feature encoder and the training objective is to reproduce the quantized $\tilde{q_t}$ latent speech representation from a set of $K$+1 distractors including candidate vector $q_t$ and $K$ distractors $\in$ Q for masked time stamp vectors at the end of contextual transformer block. The distractors are uniformly sampled from  masked frames of the same speech utterance. 
\begin{equation}
    \mathcal{L}_{cont} = -\log\frac{\exp(sim(c_t, q_t)/\tau)}{\sum_{\tilde{q}\in Q}\exp(sim(c_t, \tilde{q})/\tau)}
    \label{eq:contrastiveloss}
\end{equation}
where $sim(c_t, q_t)$ computes the cosine similarity between the quantized vector $q_t$ and contextualized transformer vector $c_t$.
\par 
The Wav2vec2.0 self-supervised model is pre-trained on 960 hours of LibriSpeech dataset and then fine tuned for ASR using CTC loss function by adding a linear layer on top of contextual block.
\par 
As the Wav2Vec2.0 showed remarkable improvement in ASR, and in stuttering speech, most parts of the speech utterance are peturbed, and it seems a plausible way to employ and explore the role of the contextual and encoder representations in SD. In this work, without fine-tuning, we employ a total of 13 contextual embeddings extracted from a local encoder and from 12 layers of the contextual transformer block as depicted in Fig.~\ref{fig:wa2vec2.0}.

\vspace{-0.3cm}

\section{Classifier description}

\subsection{K-nearest neighbourhood}
 \vspace{-0.15cm}
A non-parametric supervised algorithm based on distance metric is used mostly for classification tasks. The prediction of the query sample depends on the vote majority of its $K$ nearest neighbors~\cite{murphy2012machine}. In this work, use the Minkowski metric distance from~eq.~(\ref{eq:minsk}) with $p=2$ to fit the $K$-NN on SEP-28k dataset using embeddings computed from pre-trained ECAPA-TDNN and Wav2Vec2.0.\vspace{-0.1cm}
\begin{equation}
\small \vspace{-0.1cm}
    D = \left[ \sum_{i=1}^{k}\|x_i - y_i\|^p)\right]^{1/p}
    \label{eq:minsk}
\end{equation}
\subsection{Gaussian back-end}
 \vspace{-0.15cm}
Naive Bayes classifier (NBC) is simply a Bayesian network to handle continuous features by representing the likelihood of features using Gaussian distribution~\cite{murphy2012machine}. Given a data set $\mathcal{D} = (X_i,d_i)$ of $N$ samples with $\mathcal{R}^{1\times K}$\footnote{K is 768 for Wav2Vec2.0 and 192 for ECAPA-TDNN} pre-trained features ,~NBC assumes that the likelihood of class conditional densities is normally distributed by
\begin{equation}
    p(\bm e|C=c,\bm \mu_c, \bm\Sigma_c) = \mathcal{N}(\bm e |\bm \mu_c, \bm\Sigma_c)  
\end{equation}
where $\bm e \in X$, is extracted pre-trained representation, $\bm\mu$ and $\bm\Sigma$ are class-specific mean vector and covariance matrix respectively. The posterior probability for each target class is computed then by Bayes' formula: \vspace{-0.3cm}
\begin{equation}
\small  
    \underbrace{p(C=c|\bm e, \bm\mu_c,\bm\Sigma_c)}_{\text{class posterior}} = \frac{\overbrace{p(\bm e|C=c,\bm \mu_c, \bm\Sigma_c)}^{\text{class conditional likelihood}}p(C=c)}{\sum_{i=1}^{K} p(\bm e|C=i,\bm \mu_i, \bm\Sigma_c)p(C=i)}
    \label{eq:gaussianbackend}
\end{equation}
and then the query sample $\bm e$ is classified by taking \emph{argmax} over the classes. \vspace{-0.3cm}
\begin{equation}
    \hat{y}(\bm e) = \mathrm{argmax}~p(C=c | \bm e, \bm\mu_c, \bm\Sigma_c)
    \label{eq:argmax}  
\end{equation}
\subsection{Neural network}
 \vspace{-0.15cm}
A neural network (NN) with just few layers can also be applied on top of the pre-trained embeddings extracted from ECAPA-TDNN and Wav2Vec2.0. In this work, we use two branched NN with \emph{FluentNet} differentiating between fluent and stuttered utterances, and \emph{DisfluentNet} classifying stuttered speech utterances into several disfluency types. Each branch is composed of three FC layers with each layer followed by ReLU activation and 1D BN functions. A dropout of 0.2 is applied for first two FC layers. A Softmax layer is used in the end to get the desired classes. Following the approach in~\cite{sep28k,stutternetmtl}, for \emph{FluentNet}, we employ a pseudo labelling scheme, where we re-label all different disfluent speech samples as one class, and train the binary \emph{FluentNet} branch to differentiate between fluent and stuttered utterances. For multiclass \emph{DisfluentNet} branch, we train it by penalizing the fluent class with zero loss. During evaluation phase, the outcome from \emph{FluentNet} is considered if the prediction is fluent class, otherwise predictions from \emph{DisfluentNet} are taken into consideration.  
\section{Experimental setup}
\subsection{Embedding extractors}
\textbf{ECAPA-TDNN}: For each three second speech utterance of SEP-28k dataset, we extract speaker embeddings of dimension $\mathcal{R}^{1\times192}$ after the FC layer of ECAPA-TDNN as shown by purple line in Fig.~\ref{fig:ecapatdnn}, resulting in $\mathcal{R}^{N\times192}$ data (N is total samples). We use SpeechBrain toolkit~\cite{speechbrain} for the extraction of 192-dimensional speaker embeddings. In addition, we also use linear discriminant analysis (LDA) for dimensioanlity reduction with component size four resulting in $\mathcal{R}^{N\times4}$ dimensional data before passing it to the downstream classifiers.
\newline 

\textbf{Wav2Vec2.0}: For contextual embedding extraction of each SEP-28k sample, we extract $\mathcal{R}^{T\times 768}$ (T is temporal dimension) dimensional representations from local feature encoder $\mathcal{Z}$ and from each layer of contextual transformer block $\mathcal{C}$, each of which yields different contextual representation as shown by purple block in Fig.~\ref{fig:wa2vec2.0}. Before passing each layer representation separately to the downstream classifiers, we apply statistical pooling across temporal domain and concatenated the mean and standard deviation resulting in a feature vector of $\mathcal{R}^{1\times768\times 2}$. Moreover, we also concatenate the contextual embeddings of local encoder (L1), L7 and layer L11 of $\mathcal{C}$ after applying the statistical pooling and LDA which results in $\mathcal{R}^{N\times12}$ feature vector. We extract embeddings from the PyTorch version of Wav2Vec2.0~\cite{pytorch}.

\begin{figure*}[h]
\vspace{-1cm}
 \begin{minipage}[t]{0.33\linewidth}
    \centering
    \includegraphics[width=1\textwidth]{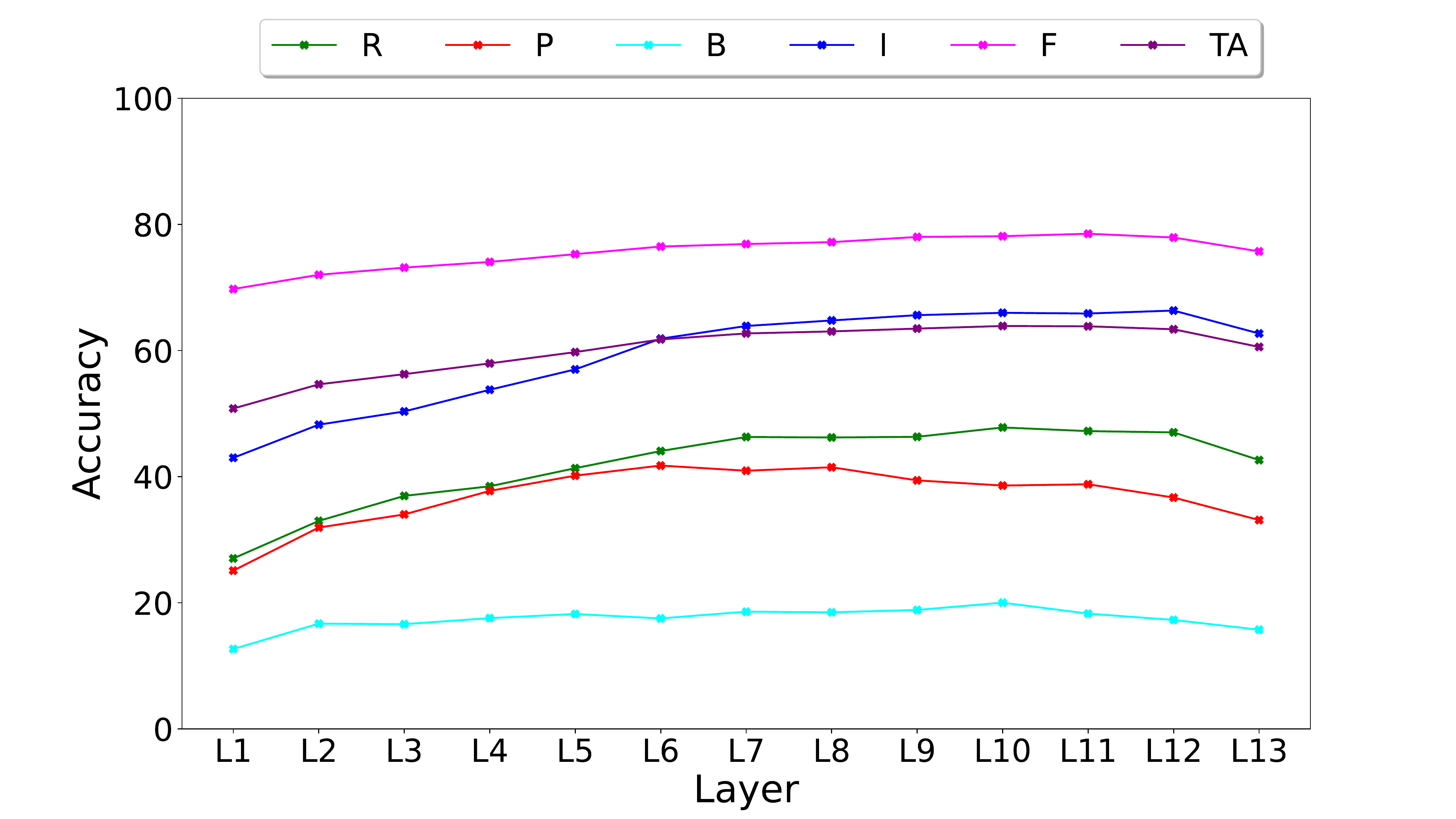}  
\end{minipage}
\begin{minipage}[t]{0.33\linewidth} 
    \centering
    \includegraphics[width=1\textwidth]{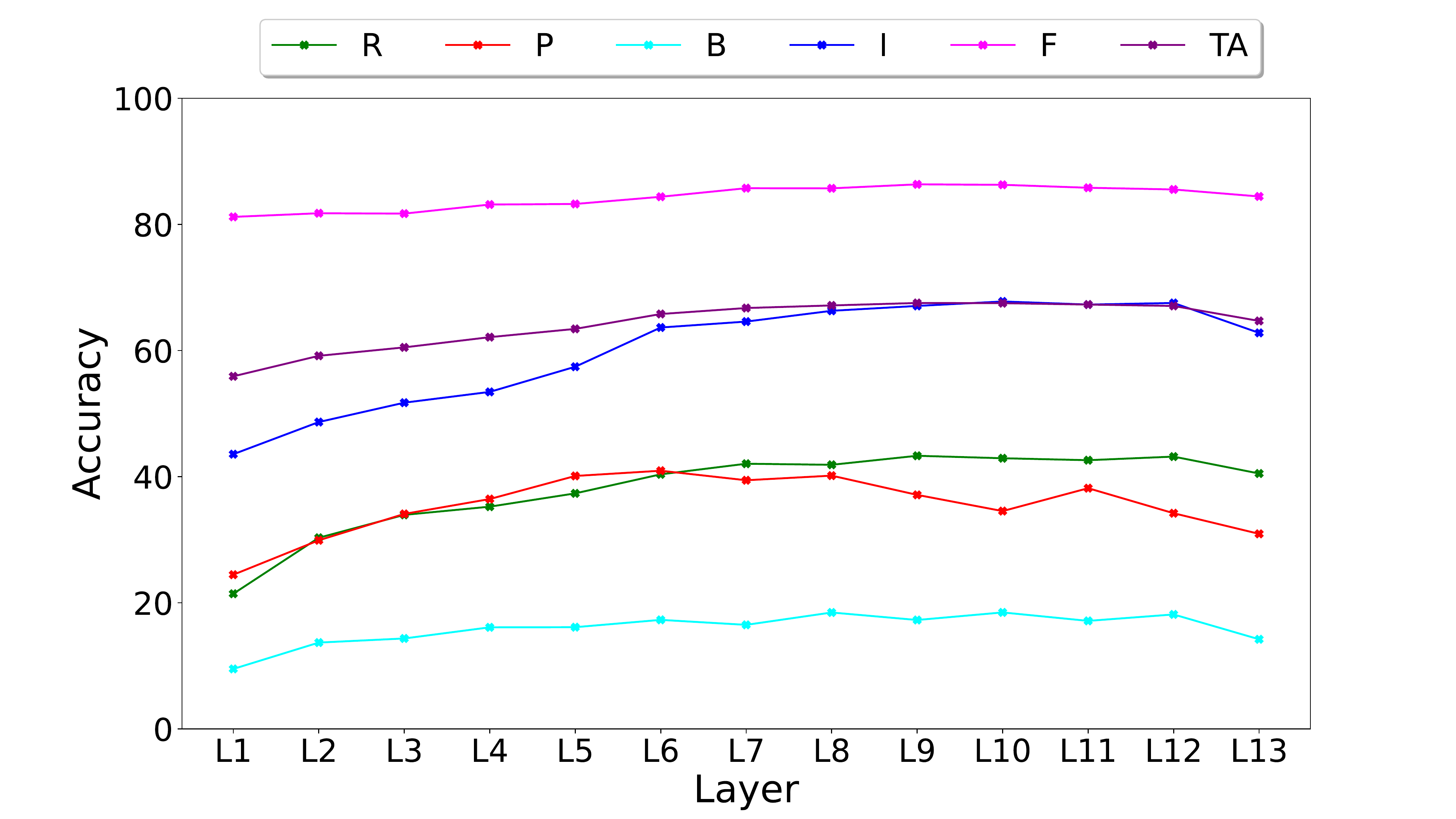}
\end{minipage} 
\begin{minipage}[t]{0.33\linewidth}
    \centering
    \includegraphics[width=0.9\textwidth]{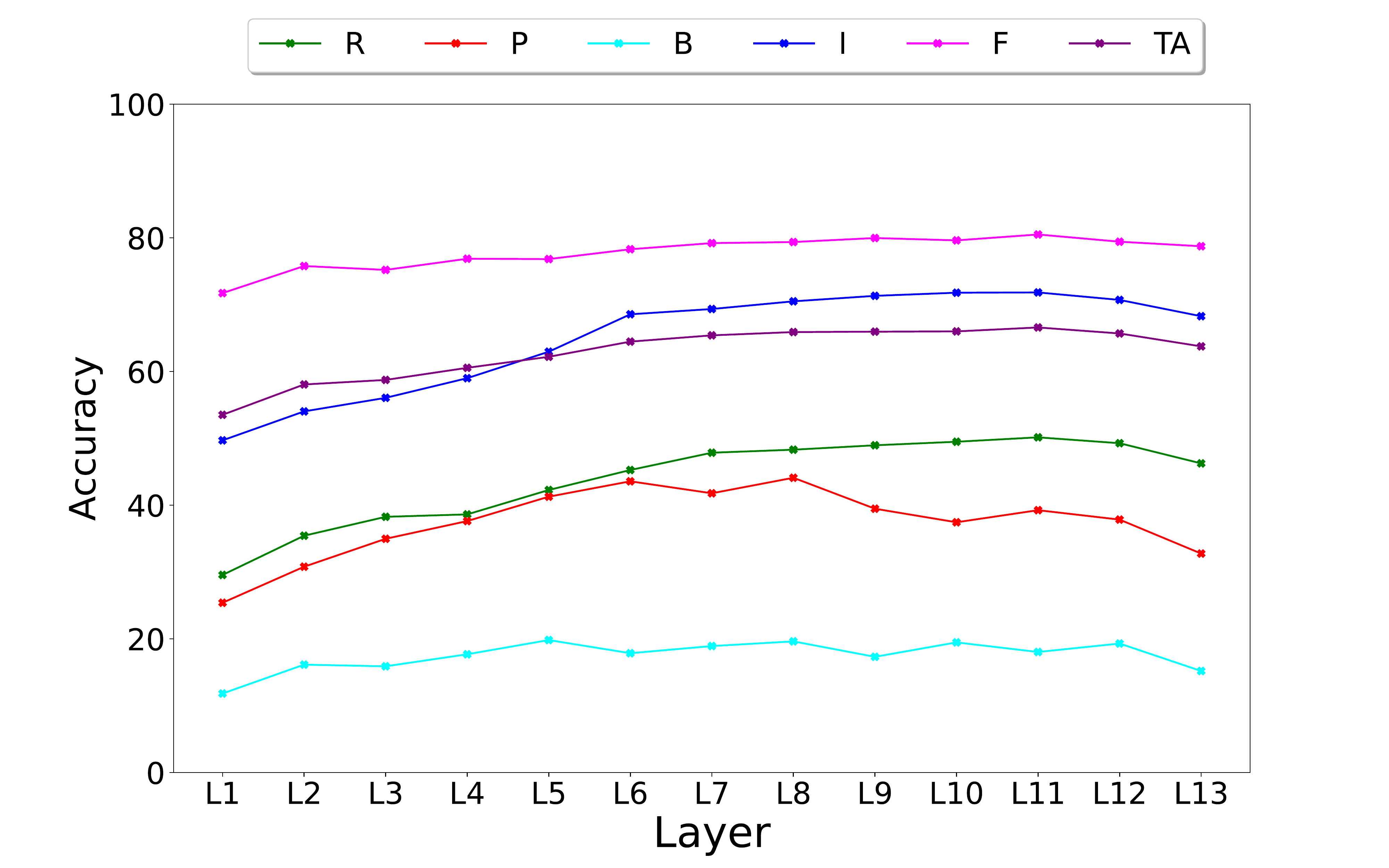}  
\end{minipage}
\vspace{-0.5cm}
\caption{Impact of various Wav2Vec2.0 contextual layers in SD with KNN (L), NBC (M), and NN (R).}
\vspace{-0.3cm}
\label{fig:layerimpact}
\end{figure*} 

\vspace{-0.2cm}
\subsection{Dataset description and evaluation metrics}
\vspace{-0.15cm}

In this case study, we used SEP-28k stuttering dataset~\cite{sep28k}  which consists of 28,177 speech samples from 385 podcasts. We have used only 23573 annotated speech segments. We randomly selected 80\% podcasts for training, 10\% podcasts for validation and the remaining 10\% of the podcast for evaluation in a 10-fold cross validation scheme. The experimental results mentioned are the average of 10-fold cross validation experiments, and are compared to the baseline results from~\cite{stutternet} which is trained on MFCC features. 
\vspace{-0.2cm}
\subsection{Implementation}
\vspace{-0.15cm}
For implementing the proposed pipeline for NN, we use PyTorch library~\cite{pytorch}, and for LDA, KNN and NBC, we have used the Scikit-learn~\cite{scikit-learn} toolkit. We have chosen value of $K=5$ in the KNN classifier. The downstream NN is trained using cross entropy optimized by the Adam optimizer with a learning rate of 1e-2, and the training is stopped using an early stopping scheme on validation loss with a patience of seven. \vspace{-0.25cm}

\section{Results and discussion }
\vspace{-0.1cm}

\begin{figure}
    \centering
     \includegraphics[width=1\columnwidth]{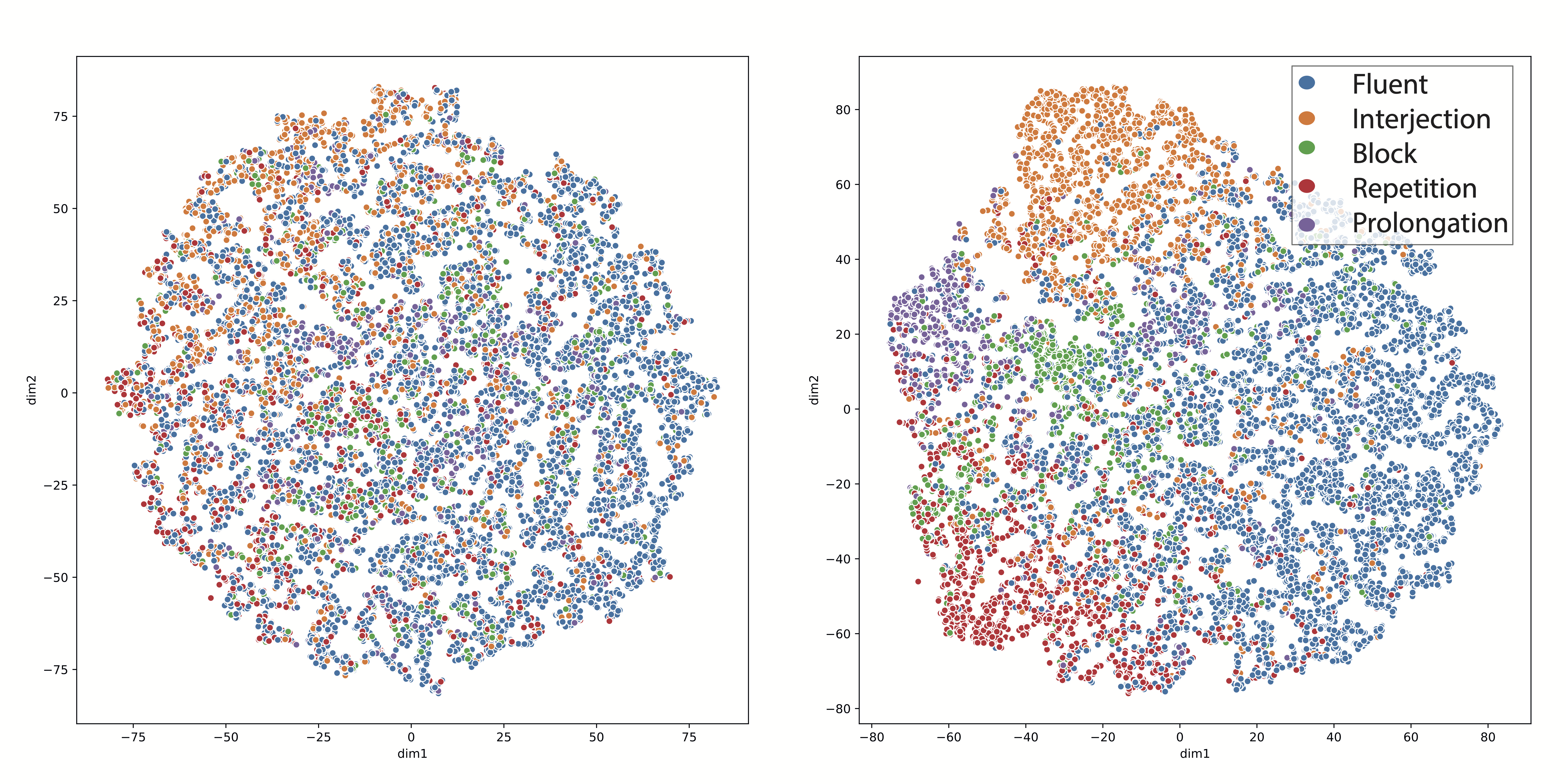}
    \caption{\small t-SNE embeddings from the FC layer of ECAPA-TDNN (L) and from L12 of contextual block $\mathcal{C}$ of Wav2Vec2.0 (R).} \vspace{-0.65cm}

    \label{fig:embed} 
\end{figure}

For thorough evaluation, we train each of the proposed models 10 times. Table~ \ref{tab:fusion} shows the average accuracy results of different stuttering and fluent classes on exploiting different features extracted from ECAPA-TDNN and Wav2Vec2.0. We compare the proposed speaker and contextual embeddings based SD models to our previous work \emph{StutterNet}~\cite{stutternet} and to the multi branched \emph{StutterNet} having two different branches with one branch differentiting between fluent and stutter utterances and the other branch distinguishing among different disfluency types. We use LDA for dimensionality reduction of features in each case before passing them to the classifiers for prediction. 
\par 
\textbf{Speaker embeddings}:
We see from the Table~\ref{tab:fusion}, that the downstream classifiers trained on ECAPA-TDNN embeddings perform poorly in all the stuttering classes as compared to the baseline results from Table~\ref{tab:fusion}. This is evident from the Fig.~\ref{fig:embed} as well, where the different stuttering type utterances are mixed and no clear cluster is visible among the disfluency classes. Furthermore, applying magnitude normalization on ECAPA-TDNN embeddings before passing them to the downstream classifiers, improves the SD performance marginally in majority of classes. The ECAPA-TDNN is trained and adapted towards speaker identification task and it is likely possible that the information (such as linguistic content, prosody, emotion state) that isn't essential for that task but could be crucial for SD gets removed from the latent embeddings. 
\par 
\textbf{Wav2Vec2.0 contextual embeddings}:
Table~\ref{tab:fusion} show the results with last but two layer of contextual transformer block $\mathcal{C}$ of Wav2Vec2.0 with and without prior application of LDA respectively. From the results, we can observe that for SD, the contextual embeddings from Wav2Vec2.0 outperforms in all the classes with an overall relative improvement over \cite{stutternet} of 5.82\% using KNN,	11.54\%	using NBC, 10.38\% using NN and over BL by 11.92\% using KNN, 17.97\% using NBC and 16.74\% using NN. Figure~\ref{fig:layerimpact} shows the impact of different contextual embeddings and local encoder representations in the detection of various stuttering classes. The plot shows almost a close trend in all the stuttering class accuracies. The detection accuracy of stuttering classes increases with the layer (L1 is local encoder, L13 is last layer of $\mathcal{C}$) number with (L) for KNN, (M) for NBC, and (R) for NN. We hypothesize that the lower layers including local encoder (L1) representations contain speech information only from the local window of size 25~ms, and, in addition passing the representations to the downstream classifiers after applying statistical pooling layer further restricts it to learn more stutter specific patterns. In addition, the results show that the contextual layers from L6 to L12 of Wav2Vec2.0 model trained in a self-supervised fashion are able to capture rich stuttering patterns as also depicted in Fig.~\ref{fig:embed}. As for the last layer (L12), it slightly degrades performance in SD in comparison to its previous layer due to the fact that the transformer block $\mathcal{C}$ was fine tuned and adapted for ASR task. By fine tuning towards ASR, it is possible that the Wav2Vec2.0 model has not focused on the information which is relevant to stuttering, resulting in loss of rich stuttering information. Consider such an example of prosodic information, which is very essential in SD, but not that important for ASR. Using Wav2Vec2.0 embeddings with NN in SD, there is an overall relative improvement of 74.74\%, 3.59\%, 88.20\%, 24.64\%, and 8.16\% in repetitions, prolongations, blocks, interjections, and fluents respectively, thus outperforms in all the cases over the state-of-the-art results. 
\par 
Moreover, the prior application of LDA on Wav2Vec2.0 representations boosts detection performance in repetitions by 13.59\%, prolongation by 14.77\%, blocks by 49.88\%, interjections by 8.24\%, and fluents by 3.90\%. This results in an overall boost by 7.28\%.
\par 
\textbf{Fusion}: In addition, we fuse the ECAPA-TDNN and Wav2Vec2.0 embeddings via score and embedding fusion schemes, the results of which can be seen from Table~\ref{tab:fusion}. While computing the final score $p$ from ECAPA-TDNN and Wav2Vec2.0 prediction probabilities, we empirically optimize the weighting parameter $\alpha$ on test set in $p = \alpha * p_{w2v2} + (1 - \alpha) * p_{ecapa}$ and we found $\alpha = 0.9$ gives the best results. The ECAPA-TDNN representations which contain rich information about speakers' identity further enhances the detection performance of fluent speech class over Wav2Vec2.0 embeddings by a relative margin of 5.34\% using KNN, 1.13\% using NBC, and 5.13\% using NN. However, it doesn't contain enough rich information about suprasegmental, emotional content, etc, which are incredibly important for disfluent classes, thus acts as a negative transfer for them. 
\par 
Each layer of Wav2Vec2.0 model contains different speech representations, exploiting this fact, we integrate the representations after applying LDA from local encoder (L1), L7 and L11 from contextual block $\mathcal{C}$, resulting in a $\mathcal{R}^{N\times12}$ dimensional data. 
Concatenating information from multiple layers improves minority class recognition including: prolongations by 8.74\% with KNN, 25\% with NBC, and 3.9\% with NN, and blocks by 14.7\% with KNN, 66.4\% with NBC, and 32.3\% when using NN.

  \begin{table}[th]
  \caption{SD results (TA: Total accuracy, B:
Block , F: Fluent , R: Repetition , P: Prolongation , I: Interjection, BL: Baseline) for different methods. The results reported for Wav2Vec2.0 are from L11.}
\vspace{-0.2cm}
  \label{tab:fusion}
  \centering
  \scalebox{0.8}{\begin{tabular}{ c c c c c c c}
    \toprule
    \multicolumn{1}{c}{\textbf{Model}} &                                     \multicolumn{1}{c}{\textbf{R}} &                                     \multicolumn{1}{c}{\textbf{P}} &   
    \multicolumn{1}{c}{\textbf{B}} & 
    \multicolumn{1}{c}{\textbf{I}} & 
    \multicolumn{1}{c}{\textbf{F}}   &
    \multicolumn{1}{c}{\textbf{TA}} \\
     \midrule
    \emph{StutterNet}~\cite{stutternet}& 21.99& 27.78 & 1.98 & 49.99 & 88.18 & 60.33\\
         BL & 28.70&	37.89	&	9.58	&	57.65	&	74.43	&	57.04\\
         \midrule
          &\multicolumn{6}{c}{\textbf{Embedding: ECAPA-TDNN}}  \\ 
\midrule
    KNN& 17.82&	6.40&	6.09&	26.58&		71.55&	45.73\\
    NBC & 22.15	&13.84&	9.42&	31.46&		63.57&	44.04\\
    NN & 23.06 &	11.67 &	5.79 &	43.29 &		69.48 &	49.12   \\
    
     \midrule
    KNN~+~LDA  &21.92&	11.93&	8.56&	26.6& 	66.77&	44.37\\
    NBC~+~LDA  &11.99&	7.24&	2.52&	25.43&	88.77&	53.53\\
    NN~+~LDA  &24.51	&10.33&	5.03&	44.49&	68.73&	48.81\\

    \midrule
    & \multicolumn{6}{c}{\textbf{Embedding: Wav2Vec2.0}}  \\     
     \midrule
        KNN  &  24.48&	8.88&	11.33&	54.02&	 84.10&	58.85 \\
    NBC  &  44.61&	5.20&14.70&	50.44&	52.16&	44.49 \\

    NN & 44.15&	34.2&	12.03&	66.36	&	77.48&	62.07 \\
\midrule

    KNN~+~LDA &47.22	&38.79&	18.28&	65.87&		78.52&	63.84\\
    NBC~+~LDA &42.61	&38.17	&17.13	&67.29		&85.81	&67.29\\
NN~+~LDA&50.15&	39.25&	18.03&	71.83&	80.50&	66.59\\
    \midrule
    \multicolumn{7}{c}{\textbf{Score fusion: ECAPA-TDNN and Wav2Vec2.0}} \\
    \midrule
       KNN &44.06	&36.95	&16.13	&64.61	&82.70	&65.18 \\
       NBC &41.75	&38.63	&16.60	&66.83	&86.78	&67.73 \\
       NN &43.58	&38.94	&19.11	&67.36	&84.63	&67.26 \\
         \midrule
          \multicolumn{7}{c}{\textbf{Embedding fusion: ECAPA-TDNN and Wav2Vec2.0}}  \\ 
         \midrule
        KNN&45.38		&37.29		&17.92		&62.18			&80.68		&64.08 \\
        NBC&44.22		&40.01		&19.70		&67.61			&84.55		&67.44  \\
         NN&44.13		&38.02		&18.39		&66.44			&84.53		&67.00  \\
         \midrule
         \multicolumn{7}{c}{\textbf{Embedding fusion: L1 + L7 + L11}}  \\ 
         \midrule
         KNN&46.98	&42.18	&20.96	&66.28		&82.24	&66.30 \\
         NBC&\textbf{48.46}	&\textbf{47.84}	&\textbf{28.51}	&\textbf{70.41}		&80.33	&67.45 \\
         NN&46.79	&40.79	&23.86	&69.54	   &\textbf{84.32}	&\textbf{68.35} \\
       
    \bottomrule
    \vspace{-0.8cm}
  \end{tabular}}
  \end{table}

 \vspace{-0.2cm}
\section{Conclusions}  \vspace{-0.05cm}

The voice pathology datasets are relatively small and expensive to collect, and stuttering is no different, thus restricts exploiting the use of advanced DL architectures for SD. 
To this bottleneck, we employ pre-trained models in SD that are trained on huge amounts of data. We exploit speaker and contextual embeddings extracted from the ECAPA-TDNN and Wav2Vec2.0 respectively. Compared to the baseline, the results indicate that the Wav2Vec2.0 contextual embeddings based SD methods achieve considerable improvement and outperform in all disfluent categories with an overall relative improvement of 10.34\% and 16.74\% over \emph{StutterNet}~\cite{stutternet} and BL respectively. 
\par 
This work can be extended by exploring fine-tuning of the Wav2Vec2.0 model for identifying and detecting the location of stuttering in speech frames. Furthermore, the generalizaion ability of the proposed method across multiple datasets is another interesting topic to study.


\section{Acknowledgements}
 \vspace{-0.1cm}
This work was made with the support of the French National Research
Agency, in the framework of the project ANR BENEPHIDIRE (18-CE36-
0008-03).

\bibliographystyle{IEEEtran}

\bibliography{ref}

\end{document}